\documentstyle[12pt,aasms4]{article}
\input epsf.sty
\newcommand{\cyg}{\mbox{Cyg X-1}}
\newcommand{\cmsec}{\rm cm^{-2}\,sec^{-1}}
\newcommand{\rms}{\mbox{r.m.s.}}
\received{1995 December 18}
\accepted{1996 March 1}
\begin{document}
\title{Correlation Between BATSE Hard X-ray Spectral and Timing Properties
of Cygnus X-1}
\author{D. J. Crary\altaffilmark{1}, C. Kouveliotou\altaffilmark{2}, 
        J. van Paradijs\altaffilmark{3,4}, 
        F. van der Hooft\altaffilmark{4}, D. M. Scott\altaffilmark{2},
        \linebreak 
        W. S. Paciesas\altaffilmark{3},
        M. van der Klis\altaffilmark{4},
        M. H. Finger\altaffilmark{2},
        B. A. Harmon\altaffilmark{5},
        and W. H. G. Lewin\altaffilmark{6}}
\altaffiltext{1}{NAS/NRC Research Associate, NASA Code ES-84, Marshall 
    Space Flight Center, Huntsville, AL\ \ 35812, USA}
\altaffiltext{2}{Universities Space Research Association, Huntsville,
    AL\ \ 35806, USA}
\altaffiltext{3}{Department of Physics, University of Alabama in Huntsville,
    Huntsville, AL\ \ 35899, USA}
\altaffiltext{4}{Astronomical Institute ``Anton Pannekoek'', University
    of Amsterdam \& Center for High-Energy Astrophysics, Kruislaan 403, 
    NL-1098 SJ Amsterdam, The Netherlands}
\altaffiltext{5}{NASA/Marshall Space Flight Center, Huntsville, AL 35812, USA}
\altaffiltext{6}{Massachusetts Institute of Technology, 37-627 Cambridge,
    MA 02139, USA}

\begin{abstract}

We have analyzed approximately 1100 days of Cygnus X-1 hard X-ray data 
obtained with BATSE to study its rapid variability. We find for the first 
time correlations between the slope of the spectrum and the hard X-ray 
intensity, and between the spectral slope and the amplitude of the rapid 
variations of the hard X-ray flux. We compare our results with 
expectations from current theories of accretion onto black holes.

\end{abstract}

\section{Introduction}

Accreting black-hole candidates (BHC) 
in X-ray binaries show three \lq source states' 
that are distinguished by characteristic spectral and (correlated) 
fast-variability properties. They are called the \lq low state', 
\lq high state' 
and \lq very high state'; the dominant parameter that determines the 
source state is likely to be the mass accretion rate (see Van der Klis
\cite{vdk:b}, and Tanaka \& Lewin \cite{Tanaka:Lewin} for recent reviews).

In the low state the X-ray spectrum is dominated by 
a very hard power law component 
extending to several hundreds of keV. The source intensity shows strong 
fluctuations, with broad-band r.m.s.\ amplitudes as high as 40 percent. 
At frequencies above $\sim\! 1$ Hz the power density spectrum 
(PDS) of these variations follows a power law; at frequencies below 
a low-frequency cut off the PDS is flat. In several BHC 
the cut-off frequency $\nu_{\rm c}$ has been observed to vary by up to an 
order of magnitude, while the high-frequency part of the PDS remained 
approximately constant (Belloni \& Hasinger \cite{Belloni}; Miyamoto et 
al. \cite{Miya:canon}). 
As a result, the power integrated over a 
frequency range below $\nu_{\rm c}$ is strongly anti-correlated 
with that cut-off frequency. 

In the high state the X-ray spectrum contains an ultra-soft thermal 
component, with (bremsstrahlung) temperatures of order 1 keV. In some 
sources (e.g., LMC~X-3, see Tanaka \cite{Tanaka:ESLAB}) the power-law 
spectral component is not 
detected, but in others (e.g., GS~1124$-$68, see 
Ebisawa et al.\ \cite{Ebisawa:Spec}) it is observed in 
combination with the ultra-soft component. In the 1--10 keV range the 
amplitude of the intensity variations is correlated with photon 
energy; this can be understood if the variability is connected with the power 
law spectral component, which is \lq diluted' by the much lower variability of 
the ultrasoft emission (Van der Klis \cite{vdk}). 

In the very high state the PDS of BHC show 3--10 Hz 
quasi-periodic oscillations and branch structure in an X-ray color-color
diagram reminiscent
of the normal-branch (NB) state of 
Z sources (Van der Klis \cite{vdk:a}), i.e., accreting neutron 
stars with magnetic fields of order 
$10^{10}\!$~G. In view of the strong arguments that the accretion rate of 
Z sources is then close to the Eddington limit it has been suggested 
that in the very high state the BHC are likewise accreting near the 
Eddington limit (Van der Klis \cite{vdk:b}).

In their low state the BHC are remarkably similar 
to atoll sources, i.e., accreting neutron stars 
with magnetic field strengths generally believed to be below $10^9\!$~G 
(substantially weaker than those of Z sources). At low accretion rates, 
i.e., in their so-called \lq island state', atoll sources have power law 
X-ray spectra (see, e.g., Barret \& Vedrenne \cite{Barret}). On 
average, these spectra 
appear to be not quite as hard as those of low-state BHC (see, e.g., 
Gilfanov et al.\ \cite{Gilfanov}); however, the distribution of 
the spectral slopes of BHC and 
atoll sources shows clear overlap (Wilson et al.\ \cite{Wilson};
Ebisawa et al.\ \cite{Ebisawa:Spec};
see also Van Paradijs \& Van der Klis \cite{jvp:vdk}). Also, the PDS of 
island-state atoll sources are very similar to those of low-state BHC. They
follow a power law at high frequencies, and are flat below a cut-off 
frequency; the high-frequency part of the PDS of the atoll source 1608$-$522 
was observed to remain approximately constant as the cut-off frequency 
varied, like for Cyg X-1 (Yoshida et al.\ \cite{Yoshida}). 

As part of our attempt to gain a better understanding of the similarities 
between accreting BHC and neutron stars we are studying the variability of 
bright BHC using the almost continuous record provided by the
Burst and Transient Source Experiment (BATSE) on the Compton Gamma Ray
Observatory. We here 
report on a variability record of \cyg, covering
approximately 1100 days, and we show 
that the amplitude of its variations are strongly correlated with the slope 
of the hard X-ray spectrum, and that correlations exist between the amplitude
of the variations and the total flux.

\section{Data Analysis}

To quantify the rapid variability, we have created PDS from
the 1.024 second time resolution large area detector (LAD) count rate
data, (the so-called DISCLA data) 
using two energy channels covering the range 20--50 and 50--100 keV.
These data were filtered to eliminate bursts, Earth 
magnetospheric events, etc., then searched for data 
segments of 512 contiguous time bins (524.288 seconds without
gaps) when the source was above the Earth's limb.  During 
outburst of the bright transients GRO~J0422+32 and GRO~J1719$-$24
the data from detectors that had these sources in their fields of view
were eliminated from consideration.
On average, we selected 40 intervals of these 512-bin data strings
per day, amounting to a total data set of $2.2 \times 10^7$ seconds.

The data for each detector and each energy channel were individually
fit to a quadratic polynomial and the fit residuals converted to
Fourier amplitudes using standard fast Fourier transform techniques.  
Using similar 524.288 second intervals obtained when the source was occulted
by the Earth, we have determined that the quadratic detrending of the raw
data yields a background (source occulted) power level that is 
flat between 0.01~Hz and the Nyquist frequency (0.488 Hz). The power
level in these background data is very close to Poissonian; using the 
normalization of Leahy et al.\ (\cite{Leahy}), in which Poisson noise
corresponds to a power density 2.0, the background power density level
equals $2.040 \pm 0.005$.  This latter value has been subtracted from
the data, so that these corrected data reflects the contribution
to the PDS from Cyg X-1 only.
Deadtime corrections are not important for the 
count rate data used here.  

The Fourier transforms of each data 
segment were then summed coherently and converted to
PDS. We made daily averages of these PDS over all segments selected
for that day.
After subtraction of the average background noise level these 
were  normalized to the squared fractional \rms\ amplitude
per unit frequency (Van der Klis \cite{vdk:b}) using the daily averaged 
detector count rates in the 20--100 keV energy band obtained
from the BATSE occultation analysis (Harmon et al.\ \cite{Harmon:occ}).
As a measure of the variability of the hard X-ray flux of Cyg~X-1 we used 
the fractional \rms\ amplitude, $f$, of the intensity fluctuations in the
frequency range \mbox{0.03--0.488 Hz}, obtained by integrating the 
power spectral density (Van der Klis \cite{vdk:b}).

\section{Results}

From their investigation of the 1--25 keV variability of Cyg~X-1
Belloni and Hasinger (1990) found that the cut-off frequency $\nu_{\rm c}$ 
in the PDS varies between 0.01 and 0.1 Hz;  since the high-frequency part 
of the PDS remains constant, the cut-off frequency is anti-correlated with 
the power level at, and below, that frequency. Crary et al.\ (\cite{Crarya}) 
have confirmed this result for the 20--100 keV range using the data 
discussed here.

We have searched for a possible correlation of the variability of Cyg~X-1,
and its hard X-ray intensity and spectral shape. The latter
quantities were obtained from the 45--140 keV X-ray light curve 
of \cyg, which has been monitored with BATSE since the 
beginning of the CGRO mission (Paciesas et al.\ \cite{Pac:cyg}). The 
45--140 keV flux, $F$, is generally between 0.07 and 
0.15 $\cmsec$, with
no prominent long term ($>$ 50 day) trends in the data evident through
most of this period.  At about Truncated
Julian Day (TJD = JD $-$ 2440000.5) 9250 (1993 September 20), 
the source flux gradually declined over a period of about
150 days to a level of $\sim 0.01\; \cmsec$.  After that, the flux rose
within 30 days, back to a level of approximately 0.1 $\cmsec$. Although it is
not possible, on the basis of BATSE observations alone, to determine
the source state of \cyg\ during these observations, the
presence of a hard component in the BATSE energy range 
indicates that Cyg X-1 was probably in the low state during most of the BATSE 
observations; during the low-flux episode between TJD 9250 and 9430
a temporary transition to the high state may have occurred. 

In Figure 1 we show a plot of $f$ versus $F$. 
In the following discussion we will distinguish the data from the low-flux
episode (indicated by squares in Fig. 1) from the remainder of the
data (indicated by asterisks).
We have included only days when $F$ was higher 
than 0.04 $\cmsec$; below this 
flux, $f$ is dominated by detector noise
due to unresolved sources in the uncollimated
LAD field of view (Crary et al.\ \cite{Crary}), and the measured value
of $f$ becomes uncertain.  
Typical errors are shown, calculated from the variance
per bin of the daily averaged power spectra propagated through the calculation
of $f$, and from the results of the occultation
analysis averaged over an entire day.  

From Figure 1 it appears that during the low-flux episode 
(TJD 9250--9430), Cyg~X-1 showed 
relatively little variability, with $f$ generally
below 12\%. These low-flux data by themselves do not show a
correlation between $f$ and $F$; similarly, the remaining points in
Figure 1, with $f$ usually in the range 10--30\%, 
do not show such correlation either. However, all points
together follow a broad, upturning band in Figure 1, which indicates
that there is a relation between the hard X-ray flux 
of Cyg~X-1 and its variability. 

If during the low-flux episode Cyg~X-1 was in the high state, this
result would suggest a dependence of the hard X-ray variability on
source state. 
Although previous observations of other black-hole candidates have indicated
that the flux in the hard tail of the energy spectrum decreases  
as the source enters a high state (this, however, is very uncertain, see
Gilfanov et al.\ \cite{Gilfanov}, and 
Tanaka \& Lewin \cite{Tanaka:Lewin}),
the hard X-ray flux alone may not
be a good marker of source state.
However, the lack of low energy (1--10 keV)
observations prevents us from drawing a firm conclusion on this.
Note that the low value of $f$ is not the result of
dilution of the variability of a hard spectral component by a less
variable soft component. 

In Figure 2 we have plotted $f$ versus the exponent, $\alpha$, of a
power law fit to the 45--140 keV spectrum 
(Paciesas et al.\ \cite{Pac:cyg}); this
figure shows that these two quantities are strongly correlated. The
data obtained during the low-flux episode are offset systematically
from the other data, showing both a softer spectrum (by about 0.4 in
$\alpha$) and smaller variability. However, the correlation between
$f$ and $\alpha$ does not reflect this systematic offset: the same
correlation is followed by each of the two groups of points
separately.

The offset of the data from the low-flux episode from the other data,
shown in both Figures 1 and 2 leads one to suspect that the hard X-ray
flux and spectral slope of Cyg~X-1 are also correlated. Figure 3,
which shows a plot of $\alpha$ versus $F$, confirms this suspicion.
Again, the two groups of data points are clearly separated in this
diagram. However, for each of the two groups of points separately the
correlation is not apparent (for the \lq high-flux' points a correlation
in the opposite sense may actually be present).

\section{Conclusions}

We have found a strong correlation between the slope of the high-energy
(20-100 keV)
X-ray spectrum of Cyg X-1 and both its high-energy X-ray flux and the
variability thereof. During most of our observations 
Cyg X-1 showed strong variability with an amplitude 
that varied in anti-correlation with
a cut-off frequency (Crary et al.\ \cite{Crarya}) similar 
to the low-state behavior in the lower-energy 
range described by Belloni and Hasinger (\cite{Belloni}). 
It is therefore likely that we encountered Cyg X-1 mainly in the low state.
During a 150 day interval beginning in
1993 September both the high-energy
X-ray flux and its variability were extremely
low; it is possible that Cyg X-1 had then entered a high state; this
remains uncertain, due to lack of low-energy coverage.

The global correlations we have found between flux, variability and
spectral hardness suggest to us that all three are determined by a
basic system parameter; the mass accretion rate is the obvious candidate. 
A variety of models have been calculated for the structure
of the accretion disks around black holes, to explain the
two-component character of their X-ray spectra, and the suspected
relation of these spectral components with accretion rate
(Liang \& Nolan \cite{Liang}, and references therein; see also 
Haardt et al.\ \cite{Haardt:etal}; Chakrabarti \& Titarchuk \cite{Chakra}).
 
Most of these models invoke a very hot medium, e.g., a corona around the inner
disk regions, that produces the hard power law component through
upscattering of low-energy photons, and a standard Shakura-Sunyaev
disk that provides the latter. Most of these models do not address
source variability.

The fast variability of Cyg X-1 and other black holes has often been
described in terms of shot noise models; the break frequency in the
PDS reflects the decay time of the shots
(Sutherland, Weisskopf, \& Kahn\ \cite{Sutherland}; 
Miyamoto \& Kitamoto \cite{Miya:cyg}; Belloni \& Hasinger \cite{Belloni}). 
However, the X-ray spectral 
properties are usually not considered in these models.

Mineshige et al.\ (\cite{Mineshige:crit}, \cite{Mineshige:disks}),
proposed that accretion disks around black 
holes are in a self-organized critical state. Their qualitative arguments, 
aimed at an understanding of the low and high states, indicate that 
relatively hard X-ray spectra go with strong variability, and relatively 
soft X-ray spectra with weak variability. 

Chakrabarti \& Titarchuk (\cite{Chakra}) recently argued that accretion disks 
around black holes contain a shock, at several tens of Schwarzschild 
radii.  In the low state, the post-shock region is quite hot, and the emergent
spectrum is very hard ($\alpha\! <\! 2.5$).  In this model, $\alpha$
increases with disk accretion rate in the low state.
Also in this context, Molteni, Sponholz, 
\& Chakrabarti\ (\cite{Molteni}) found that over a range in
mass accretion rates the cooling time of the flow inside the
shock is comparable to the infall time scale. Under these conditions
the location of the shock, and the X-ray luminosity, undergoes 
quasi-periodic oscillations.  The centroid frequency of these
oscillations increases with the mass accretion rate,
with typical values around 5 Hz for a $5 M_\odot$
black hole. 

This result may be related to the anti-correlation between $\alpha$ and 
$f$ for Cyg X-1, by using the recent results of Van der Hooft 
et al.\ (\cite{vdh}) on the PDS of the black-hole transient GRO J1719$-$24. 
They found that the detailed 
shape of the PDS remained invariant under a 
frequency shift marked by the variation of a strong QPO peak (between $\sim\! 
40$ and $\sim\! 300$ mHz), while also the power in a given (stretched and 
squeezed) \lq rest frequency' interval remained constant. 

This would suggest that the QPO frequency is proportional to a break 
frequency in the PDS of this source, and that the latter may be used as 
a frequency scaler as well. If we are allowed to generalize this result for 
GRO J1719$-$24 to Cyg X-1, at least the direction of the correlation 
between $\alpha$ and $f$ found by us for Cyg X-1 would be accounted for by 
the model described by Chakrabarti \& Titarchuk (\cite{Chakra}) 
and Molteni et al.\ (\cite{Molteni}).

At super-Eddington disk accretion rates the model of 
Chakrabarti \& Titarchuk (\cite{Chakra}) predicts that
the flow inside the shock is cooled
by Comptonization of low energy photons from the disk, and the
post-shock flow becomes predominantly radial (and converging).
Comptonization in this region can still occur due to bulk
motions, producing a hard tail with $\alpha\! \sim\! 2.5$. 
The value of $\alpha$ increases weakly with the accretion rate; note 
however that in this model 
the X-ray luminosity may not be a good measure of the accretion rate as 
part of the internal energy in the disk is advected into the black hole
(see also Narayan, McClintock, \& Yi\ \cite{Narayan}).  In this
regime the flux variability is determined by the variability of the
illumination geometry of the converging inflow, and not by the
change in the size of the post-shock region, as it is in the low state.
The amplitude variation of the hard flux is expected to be 
smaller in this case (corresponding to the high state, since the  
increase in disk accretion rate leads to a increase in soft emission)
than in the low state.  It appears, then, that these theoretical results
are in qualitative agreement with the data from the low-flux episode
that we have observed in Cyg X-1.

The correlations we have found for Cyg~X-1 would easily have escaped 
attention were it not for the 1100 days of 
continuous coverage provided by the BATSE 
all-sky monitoring capabilities. We are looking forward to an improved 
understanding of our results in terms of the source state framework for 
black holes (Van der Klis \cite{vdk}) by combining the BATSE hard X-ray 
monitoring with that provided in the near future at low energies by XTE.

\acknowledgments

We would like to thank the referee, Lev Titarchuk, for helpful comments.
This project was performed within NASA grant NAG5-2560 and supported
in part by the Netherlands Organization for Scientific Research (NWO)
under grant \mbox{PGS 78-277.} This work was performed while DJC held
a National Research Council-NASA Research Associateship.
FvdH acknowledges support by the Netherlands Foundation for Research in
Astronomy with financial aid from NWO under contract number
\mbox{782-376-011}, and
the Leids Kerkhoven--Bosscha Fonds for a travel grant.
JvP acknowledges support from NASA grant \mbox{NAG5-2755}.
WHGL acknowledges support from NASA grant \mbox{NAG8-216}.

\pagebreak

\noindent Fig. 1 -- \cyg\ fractional \rms\ amplitude (0.03--0.488 Hz)
in the 20--100 keV band versus 45--140 keV flux ($\cmsec$).

\noindent Fig. 2 -- Fractional \rms\ amplitude (0.03--0.488 Hz) in the
20--100 keV band versus photon power law index from a fit in the
45--140 keV range.

\noindent Fig. 3 -- \cyg\ 45--140 keV flux versus photon power law index
from a fit in the 45--140 keV range.

\pagebreak

\pagestyle{empty}
\begin{figure}
\epsfysize=350pt
\vspace{-1in}
\epsffile{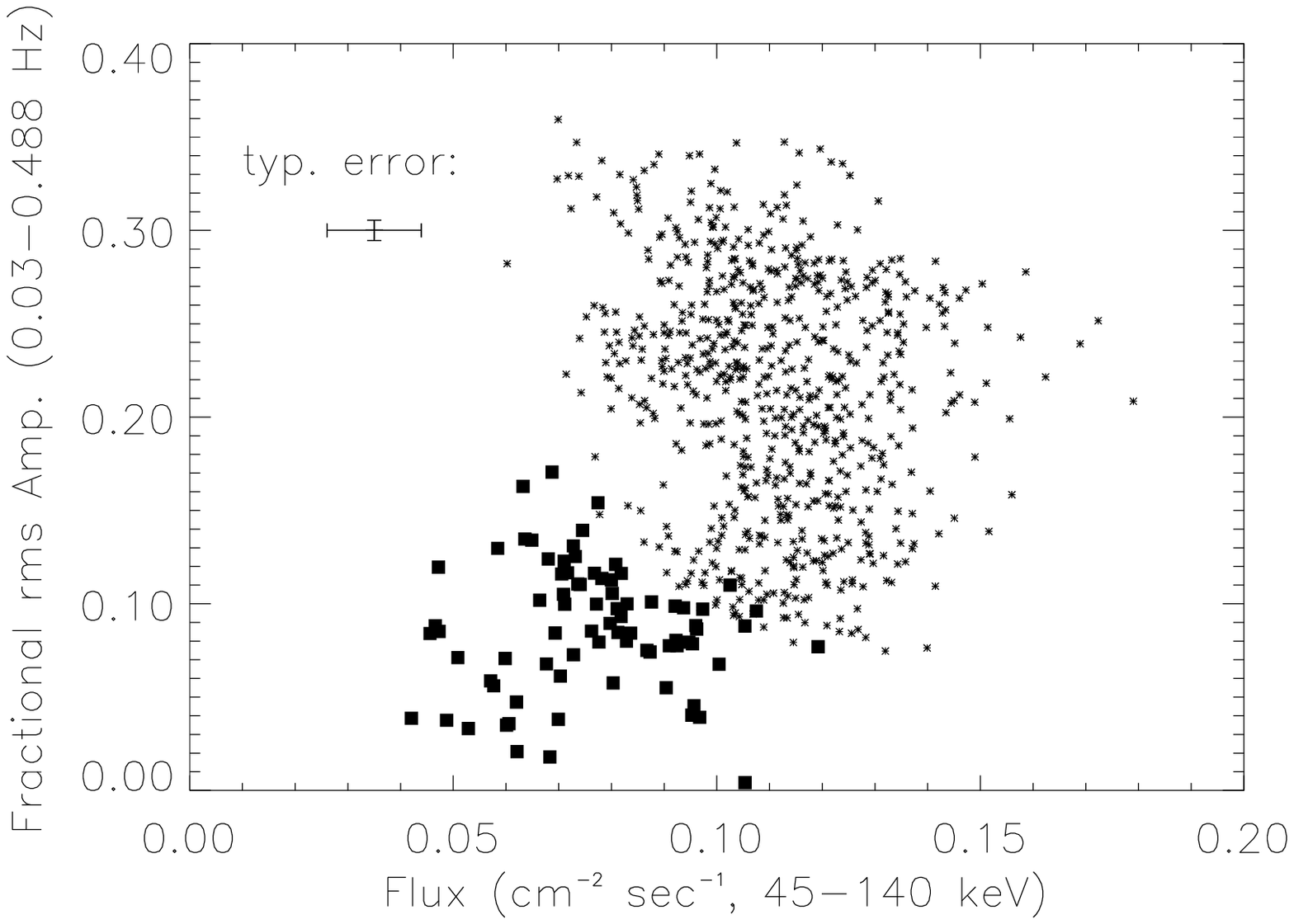}
\caption{}
\end{figure}

\pagestyle{empty}
\begin{figure}
\epsfysize=350pt
\vspace{-1in}
\epsffile{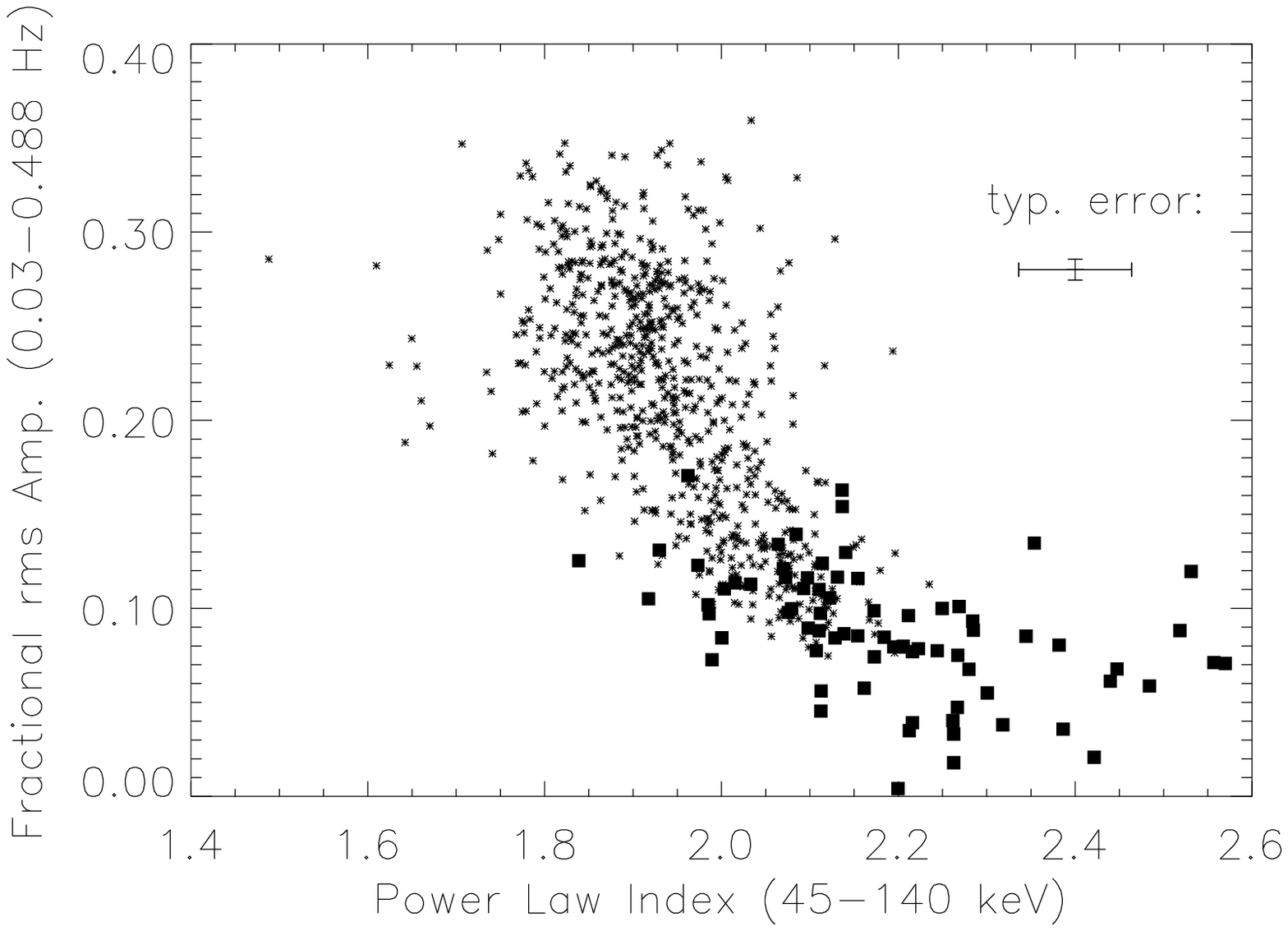}
\caption{}
\end{figure}

\pagestyle{empty}
\begin{figure}
\epsfysize=350pt
\vspace{-1in}
\epsffile{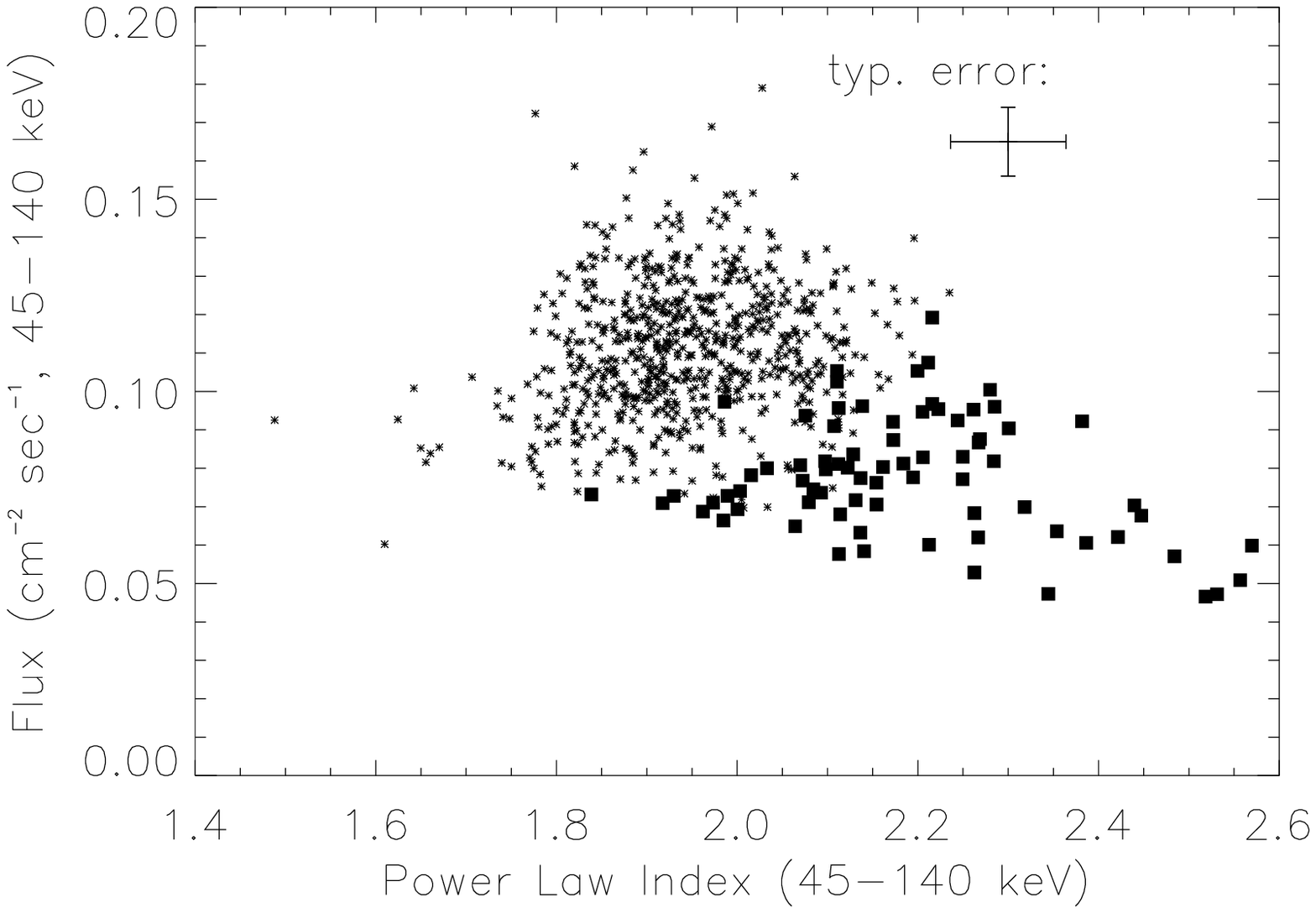}
\caption{}
\end{figure}

\end{document}